\documentstyle[aps,epsf,amsbsy,12pt]{revtex}
\begin{document}

\renewcommand{\baselinestretch}{1.66}

\title{Chiral Fluid Dynamics and Collapse of Vacuum Bubbles}

\author{I.N.~Mishustin$^{a,b}$, 
O.~Scavenius$^a$\\
$^a$ The Niels Bohr Institute, Blegdamsvej 17, DK-2100 Copenhagen {\O},
Denmark\\
$^b$ The Kurchatov Institute, Russian Research Center, 123182 Moscow, 
Russia \\
}

\maketitle

\begin{abstract}

We study the expansion dynamics of a quark-antiquark plasma droplet from an 
initial state with restored chiral symmetry. The calculations are made within 
the linear $\sigma$ model scaled with an additional scalar field 
representing the gluon condensate. We solve numerically 
the classical equations of motion for the meson fields coupled to the 
fluid-dynamical equations for the plasma.
Strong space-time oscillations of the meson fields are observed in the
 course of the chiral transition. A new phenomenon, the formation and 
collapse of vacuum bubbles, is predicted. 
The particle production due to the bremsstrahlung of the meson fields 
is estimated. \\
  \\

\end{abstract}

{\it Introduction:} It is commonly believed that the
conditions for chiral symmetry restoration and colour deconfinement
 can be reached in the course of an
ultra-relativistic heavy-ion collision. The Quark-Gluon Plasma (QGP)
is expected to be formed at some intermediate stage of the reaction. 
Since strong collective expansion may develop already in the QGP, 
its subsequent transition to the hadronic phase should be
treated dynamically \cite{Mis0}. At present this is possible
only on the basis of effective models obeying
the symmetry properties of QCD.

Our considerations below are based on the  linear $\sigma$ model 
which respects approximate chiral symmetry.
In addition to the usual chiral fields, 
$\sigma$ and $\boldsymbol{\pi}=\{\pi_1,\pi_2,\pi_3\}$, 
the model includes the dilaton or glueball
field, $\chi$, to simulate the trace anomaly of QCD \cite{Gomm}. 
The $\sigma$ and $\chi$ represent the quark and gluon
condensates as effective meson fields. 
Models of this kind were used earlier for nuclear matter 
(see e.g. \cite{Mishustin}). 

{\it The dynamical model:} The effective chiral Lagrangian for 
constituent quarks interacting with the background meson fields 
is written as
\begin{displaymath}
{\cal L}=\bar{q}\left[i\gamma_{\mu}\partial^{\mu}-
g\left(\sigma+i\gamma_5\boldsymbol{\tau}\cdot \boldsymbol{\pi}\right)
\right]q
+\frac{1}{2}
\left[\partial_{\mu}\sigma \partial^{\mu}\sigma+
\partial_{\mu}\boldsymbol{\pi}\cdot\partial^{\mu}\boldsymbol{\pi}
\right]
\end{displaymath}
\begin{equation}
-U(\sigma,\boldsymbol{\pi},\chi) 
+\frac{1}{2}\partial_{\mu}\chi\partial^{\mu}\chi
-W(\chi)-J\chi^2 .
\label{la1}
\end{equation}
\begin{displaymath}
U(\sigma,\boldsymbol{\pi},\chi)=\frac{\lambda_1^2}{4}
\left[\sigma^2+\boldsymbol{\pi}^2-\sigma_0^2
\left(\frac{\chi}{\chi_0}\right)^2\right]^2-
f_{\pi}m_{\pi}^2\sigma\left(\frac{\chi}{\chi_0}
\right)^{n},
\end{displaymath}
\begin{equation}
 ~ W(\chi)=\frac{\lambda_2^2}{4}\chi^4\ln{\left(\frac{\chi^4}
{\Lambda^4}\right)}. 
\label{la}
\end{equation}
Here $U$ is the usual Mexican Hat potential scaled by the glueball 
field $\chi$ (below we take $n=3$) and
$W$ is the effective potential responsible for the scale invariance breaking. 
The trace of the energy-momentum tensor for the above Lagrangian, 
$T_{\mu}^{\mu}=-\lambda_2^2\chi^4$, 
is assumed to be proportional to the gluon condensate,
$\langle G_{\mu\nu}^2\rangle$. The scale parameter $\Lambda$ is of the 
order of $\Lambda_{QCD}\approx 200$ MeV.

This Lagrangian leads to the normal vacuum state
where chiral symmetry is spontaneously broken: 
$\sigma=f_{\pi}=93$ MeV, $\boldsymbol{\pi}$=0, $\chi=\chi_0=136$ MeV.
The parameters of the Lagrangian are chosen so that in the normal
vacuum the constituent quark mass $m_q=gf_{\pi}=313$MeV, 
the $\sigma$-meson mass
$m_{\sigma}^2=2\lambda_1^2f_{\pi}^2+m_{\pi}^2\approx (0.6$ GeV)$^2$ and the 
glueball mass 
$m_G^2=4\lambda_2^2\chi_0^2+{\cal O}(m_{\pi}^2)\approx (1.7$ GeV$)^{2}$.
The energy density associated with breaking the gluon condensate,
$B=\lambda_2^2\chi_0^4/4$ is fixed to $0.5$ GeV/fm$^3$ \cite{SVZ}.  
In the case of thermal
equilibrium and a frozen $\chi$ field, $\chi=\chi_{0}$, 
the model leads to a chiral transition at temperatures 
around 130 MeV \cite{Mocsy}.
 
 The form of the effective glueball potential eq. (\ref{la1}) is motivated also
 by the instanton liquid model (see the recent review \cite{Shuryak}) 
if $\chi^{4}$ is identified with the instanton density.
 As predicted by this model, the instanton density is significantly
 suppressed at high temperatures. In our calculations the coupling of
 the gluon condensate to the thermal bath is parametrized in a simple
 form $J(x)\chi^{2}(x)$ (the last term in eq. (\ref{la})), where 
 $J(x)=AT^{2}(x)$ and $T(x)$ is the local temperature. The coupling
 strength $A\approx 2.4$ is chosen so that the gluon condensate,
 $\chi^{4}$, is reduced by about 60\% at $T=280$ MeV.

The model presented above is not fully consistent since it does not include exsplicitely the thermal gluon excitations. In doing so we were motivated by the recent analysis \cite{heinz} of lattice data demonstrating that at temperatures
of interest here the gluons have a rather large effective mass of about 0.6-0.8
GeV. Therefore, the contribution of thermal gluons to all thermodynamical quantities is significantly suppressed by the Boltzman factor. On the other hand, their contribution is included implicitely in the coupling term ${\cal J}\chi^2$ determining the degree of reduction of the gluon condensate at high temperatures.

Below we adopt the mean field approximation considering $\sigma$,
 $\boldsymbol{\pi}$
and $\chi$ as classical fields. The equations of motion for these fields are
obtained by applying the variational principle to the above Lagrangian. The
source terms in these equations are determined by the distribution of quarks
and antiquarks, which in principle should be found by solving the Dirac 
equation. Due to the interaction with meson fields,
quarks acquire an effective mass \cite{Csernai}
\begin{equation}
m_q(x)=g\sqrt{\sigma^2(x)+\boldsymbol{\pi}^2(x)},
\label{mass}
\end{equation}
which, in general, is space and time dependent. This makes an exact solution of
the Dirac equation very difficult. To avoid this problem one should make
further approximations. A reasonable starting point is the
 Vlasov-type kinetic equation for the scalar part of the
 quark-antiquark Wigner function $f(x,p)$,
\begin{equation}
\left[p^{\mu}\frac{\partial}{\partial x^{\mu}}+
\frac{1}{2}\frac{\partial m_{q}^{2}(x)}
{\partial x^{\mu}}\frac{\partial}{\partial p^{\mu}}\right]f(x,p)=I_{coll}[f],
\label{vla}
\end{equation}
where $I_{coll}$ is the collision term.
In refs. \cite{Csernai,MS,Abada} the collisionless ($I_{coll}=0$)
 approximation was used for the propagation of quarks in background 
meson fields. Obviously this approximation can be justified only for the late
stages of the expansion. 

Here we consider another approximation which is more appropriate
 to high temperatures. Namely, we assume
that the partonic collisions are frequent enough to maintain local
thermodynamical equilibrium. In this case $f(x,p)$ can be represented 
in terms of the equilibrium distribution functions characterized 
by local temperature $T(x)$ and chemical potential $\mu(x)$.

By multiplying both sides of eq. (\ref{vla}) with $p^{\mu}$, projecting
 on the mass shell , $p^{\mu}p_{\mu}=m^{2}(x)$, and integrating
 over 4-momenta one arrives at the equations of relativistic hydrodynamics
 \cite{MPS} 
(see also \cite{Ivanov})
\begin{equation}
\frac{\partial}{\partial x^{\mu}}T^{\mu\nu}(x)+\rho_{s}(x)
\frac{\partial}{\partial x_{\nu}}m_{q}(x)=0,
\end{equation}
where $T^{\mu\nu}(x)$ is the energy-momentum tensor,
 $T^{\mu\nu}=({\cal E}+P)u^{\mu}u^{\nu}-Pg^{\mu\nu}$,
 and  $u^{\nu}(x)$ is the collective 4-velocity of the quark-antiquark fluid. 
Here the energy density ${\cal E}$, pressure $P$ and scalar density
 $\rho_{s}$ are functions of $T(x)$, $\mu(x)$ and $m_{q}(x)$.
 They can be expressed in the standard way through the
 Fermi-Dirac occupation numbers. We solve these equations
 consistently with the equations of motion for the meson
 fields which determine the quark effective mass through
 eq. (\ref{mass}). This is why we call this approach
 Chiral Fluid Dynamics (CFD).

The evolution of the glueball field is driven by the couplings
 to the chiral fields and to the thermal bath. The corresponding
 source $J(x)$ drops with the 
characteristic hydrodynamical time of order of a few fm/c. Due to some
uncertainties in the glueball Lagrangian, particularly,
 in the derivative terms, below we consider two options of the model
, i.e. full dynamics as described above and pure chiral dynamics with a frozen 
glueball field $(\chi\equiv\chi_{0})$.


{\it Numerical results:} For numerical simulations we have used the
 RHLLE algorithm described and tested in \cite{Bernard} for fluid dynamics and the
 staggered leapfrog method for the field equations.
 Below we present results for the real-time evolution of 
spherical droplets of radii $R=2$ fm and $R=4$ fm. 
In the initial state we take a baryon-free ($\mu=0$) fluid with a 
Woods-Saxon temperature profile and a linear
profile of the collective momentum density. Initially the system is
 assumed to be in the chiral-symmetric phase at temperature
 $T\approx 280$ MeV. The initial
conditions for the fields are chosen uniformly within the droplet
 and smoothly interpolated to their vacuum values outside the droplet. 
We assume that $\sigma$ and $\chi$ fields are initially close to their
 equilibrium values at this high temperature. 

The results of the calculations for $R=4$ fm are presented 
in Figs. 1 and 2. It is seen that within a few fm/c the
 energy density of the fluid drops from the initial value
 of about 5.0 GeV/fm$^3$ to below 0.1 GeV/fm$^3$. A shell-like
 structure of the matter distribution is clearly seen at late 
times \cite{Aichelin,Rischke}. The fluid is cooled down to $T=130$ MeV
 already at $t\approx$5 fm/c. As Fig. 2 shows, at this time
 the $\sigma$ field changes rapidly from its initial value,
 $\sigma \approx 0$, towards the new asymptotic value,
 $\sigma=f_{\pi}$. This transition is accompanied by
 strong nonlinear oscillations of the all coupled fields. 

The pion field oscillations are especially strong and spread over 
the whole space within the light cone. In accordance with previous 
studies  \cite{Amelino,Wilczek,Gavin,Huang,Randrup},
 our calculations show that soft pion modes
 are indeed strongly amplified (by a factor 10) 
in the course of the chiral transition even in a finite expanding droplet. 
As suggested earlier (see e.g. \cite{Anselm,BlaiKrzy,Bjorken}),
a perfect isospin alignment of the classical pion field should
lead to a non-statistical distribution of the ratio of neutral to 
charged pions. One should however bear in mind that these coherent pions will 
be accompanied by a large number of genuine pions (in the considered example, about 
1000) produced at the hadronization of plasma.


It is interesting to note that the heavy $\sigma$ and $\chi$ fields have
quite different dynamics compared to the pion field. 
Initially they evolve almost adiabatically following
the instantaneous temperature. Instead of expanding they first 
shrink and then rebound at about the time of the chiral transition,
 when strong oscillations start. 
One can understand this behaviour from the following consideration.
The regions where $\sigma$ and
$\chi$ significantly deviate from their vacuum values, $f_{\pi}$ and $\chi_0$,
have an extra energy density $\Delta{\cal E}_{vac}\sim B$, compared to the 
normal vacuum. This vacuum energy excess generates a negative pressure,
$P_{vac}\equiv -\Delta{\cal E}_{vac}$. Such regions 
can survive only untill the internal pressure of matter 
(in our case, quark-antiquark fluid), $P_{mat}$, is large enough to 
counterbalance the external vacuum pressure. i.e. when $P_{mat}+P_{vac}\geq 0$.
This condition is always fulfilled in an equilibrated system, Howevere, in 
the course of a rapid expansion the opposite condition, $P_{mat}+P_{vac}<0$, 
can eventually be reached in a certain region of space which we call a vacuum 
bubble. Then the outside vacuum propagates into this bubble trying to minimize 
its size. This process looks like a collapse of an air bubble in a liquid. In 
our case the role of a liquid is played by the vacuum quark and gluon 
condensates. 

The collapse starts from the surface of the quark-antiquark droplet. As the
energy density of the fluid decreases the speed of the ingoing wave increases.
Finally the true vacuum penetrates to the center. Due to the inertial forces
the condensates overshoot their equilibrium vacuum values.
This is why very strong oscillations are developed at the center of the 
bubble. This violent dynamics may lead to very interesting phenomena like 
particle production by the bremsstrahlung mechanism, reheating of the fluid 
or trapping of some quarks and antiquarks (especially heavy ones $s,
\overline{s},c,\overline{c}$) in the bubble.
In this paper we consider only the first process.  

{\it Particle production:} In general, a time-dependent meson
 field $\phi({\bf r},t)$ can be represented asymptotically 
as an ensemble of quanta of this field. Using the coherent 
state formalism one can write the explicit expression for the momentum 
distribution of the mesons produced (see e.g. refs. \cite{Abada,Amelino})
\begin{equation}
2\omega_{{\bf k}}\frac{dN_{\phi}}{d^3k}=
\frac{1}{\left(2\pi\right)^3}\left[\mid \dot{\phi}({\bf k},t)\mid^2
+\omega_{{\bf k}}^2\mid\phi({\bf k},t)\mid^2\right],
\end{equation}
where $\omega_{{\bf k}}=\sqrt{{\bf k}^2+m_{\phi}^2}$ is the single-particle
 energy of a meson with the vacuum mass $m_{\phi}$. In this
formula $\phi({\bf k},t)$ and $\dot{\phi}({\bf k},t)$ are the 3-dimensional 
Fourier transforms of the meson field $\phi({\bf r},t)$ and its time 
derivative $\dot{\phi}({\bf r},t)$. They are obtained from the dynamical
simulations described above. The r.h.s. should be calculated at sufficiently
late times when nonlinearities in the field equations are negligible. 


The numbers of produced particles depends sensitively on the initial 
conditions for the fields, droplet size and the expansion dynamics. 
In Table 1 we present  the particle numbers (calculated at $t=30$ fm/c)
for the illustrative example of Fig. 2, $R=4$ fm, with the initial glueball 
field 0.79$\chi_0$ (case I) and $0.1\chi_{0}$ (case II), as well as for a 
smaller droplet, $R=2$ fm. For comparison, the results for the frozen glueball 
field, $\chi\equiv\chi_{0}$, are also shown.
The inclusion of a dynamical glueball field leads to an increase 
in the pion number by a factor of 2 or more, mainly due to the additional
 amplification of the ``pion laser'' modes \cite{Abada}. They originate
 from the splitting of soft $\sigma$ modes into pion modes with momenta 
of about $m_{\sigma}/2$. 
But a larger fraction of energy, associated with the initially suppressed  
gluon condensate, goes into the bremsstrahlung of $\sigma$ and $\chi$ fields.
One can see that much more particles, especially glueballs, are produced when 
the system is initially trapped in a metastable state with $\chi\approx 0$ 
(case II).
On the other hand, significantly fewer particles are produced
from a smaller droplet with $R=2$ fm.

{\it Conclusions:} It is demonstrated that the chiral
transition in an expanding finite droplet is accompanied by strong space-time 
oscillations of the background fields. 
The long wave-length modes of the pion field are strongly amplified,
 by a factor 10, in the course of transition. The gluon condensate
 brings into play a new scale of energy density, $B=0.5$GeV/fm$^3$,
 which changes significantly the dynamics.

The simulations reveal a novel phenomenon, the formation and collapse
of the vacuum bubbles, associated with the regions of out-of-equilibrium 
quark and gluon condensates and low matter pressure.
The additional energy released in the collapse goes partly into the
coherent pions, but to a larger extent, to the production of $\sigma$-mesons 
and glueballs. Due to a very large width of the 
$\sigma$ meson its direct observation in heavy ion collisions is practically
impossible. But the glueballs can be detected by the characteristic decay
channels  $G\rightarrow \pi\pi, \overline{K}K$ with widths of about 100 MeV,
 as well as by the electromagnetic decay ($G\rightarrow \gamma\gamma$) with 
a width of a few keV.

In the future we are planning to improve the model in two directions.
First, a more realistic treatment of the gluon condensate and its coupling
to the partonic plasma should be introduced. Second, friction terms due to 
the interaction of meson fields with the thermal bath should be included.

{\it Acknowledgements:} The authors thank J. P. Bondorf, A. Dumitru, P. Ellis, 
A. D. Jackson, D. Diakonov, D. Rischke, J. Randrup, L.M. Satarov, E. 
D. Shuryak, J. Wambach and X-N. Wang for fruitful discussions. 
This work was partly supported by the Carlsberg Foundation, Denmark.

\newpage

\begin{table}
\begin{center}
\caption{The numbers of neutral pions $(N_{\pi_{0}})$, $\sigma$ mesons
 $(N_{\sigma})$ and glueballs $(N_{G})$ produced by the brensstrahlung 
mechanism in the course of a quark-antiquark droplet expansion. 
Results are presented for initial droplet radii 2 and 4 fm, as well as 
for full (cases I and II) and frozen $\chi$ dynamics.} 

\begin{tabular}{|c|c|c|c|c|} \hline  
R  & $\chi$-dynamics & $N_{\pi_{0}}$ & $N_{\sigma}$ & $N_{G}$ \\\hline  
2 fm & frozen & 0.4 & 0.4 & 0 \\\cline{2-5}
   & full (I) & 0.4 & 0.7 & 0.5 \\\hline 
   & full (II) & 3 & 2 & 0.7 \\\hline   
4 fm & frozen & 5 & 2 & 0 \\\cline{2-5}
 & full (I) & 4 & 6 & 3 \\\hline  
 & full (II) & 13 & 10 & 17 \\\hline
\end{tabular} 
\end{center}
\end{table}

\begin{figure}
\begin{center} 
\epsfxsize=14cm
\epsffile{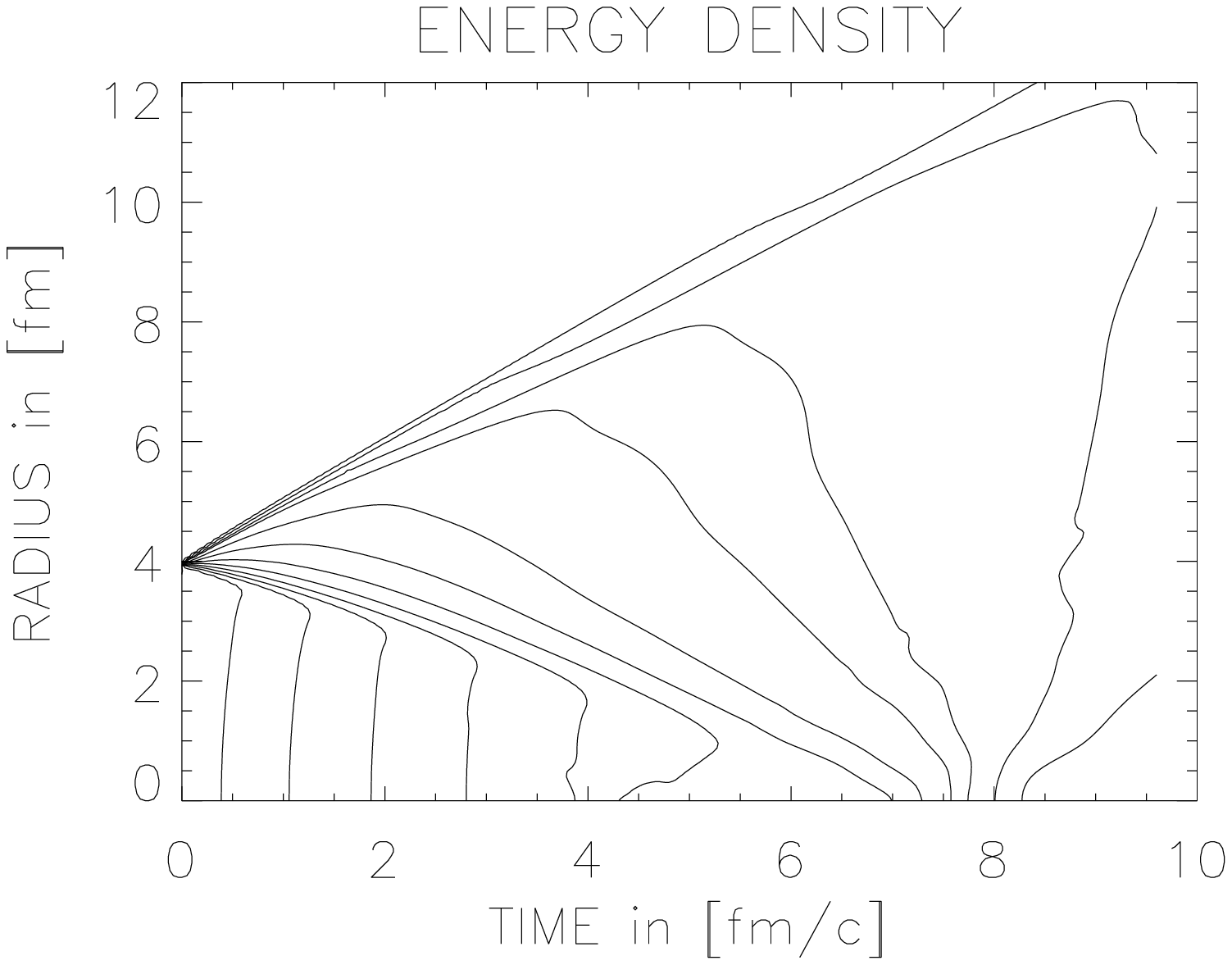}
\epsfxsize=14cm
\epsffile{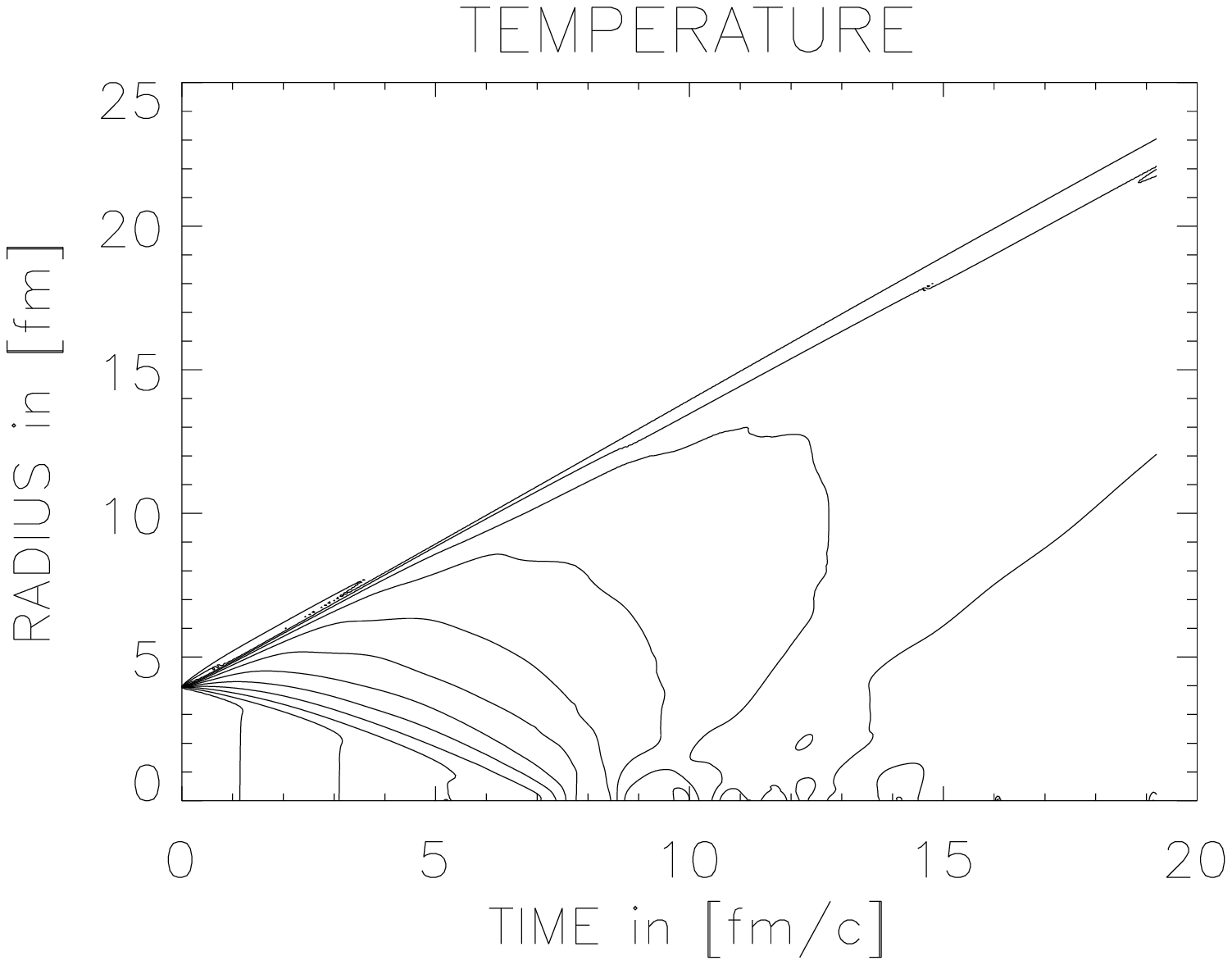}
\caption{Contour plots in the $r-t$ plane (case I) of the energy density
 in GeV/fm$^{3}$ (top from left the subsequent decreasing levels are: 5.0,4.55,4.05,3.55,3.05,1.55,2.05,2.55,1.05,0.55,0.35,0.15 and 0.05 GeV/fm$^{3}$)
and temperature in MeV (bottom, from left the subsequent decreasing levels are: 280,260,240,220,200,180,160,140,120,100,80,60 and 40 MeV)
 in the expanding spherical quark-antiquark droplet
of initial radius $R=4$ fm. The initial collective velocity is about $0.2$c at the
surface.}
\end{center}
\end{figure}

\begin{figure}
\begin{center} 
\epsfxsize=15cm
\epsffile{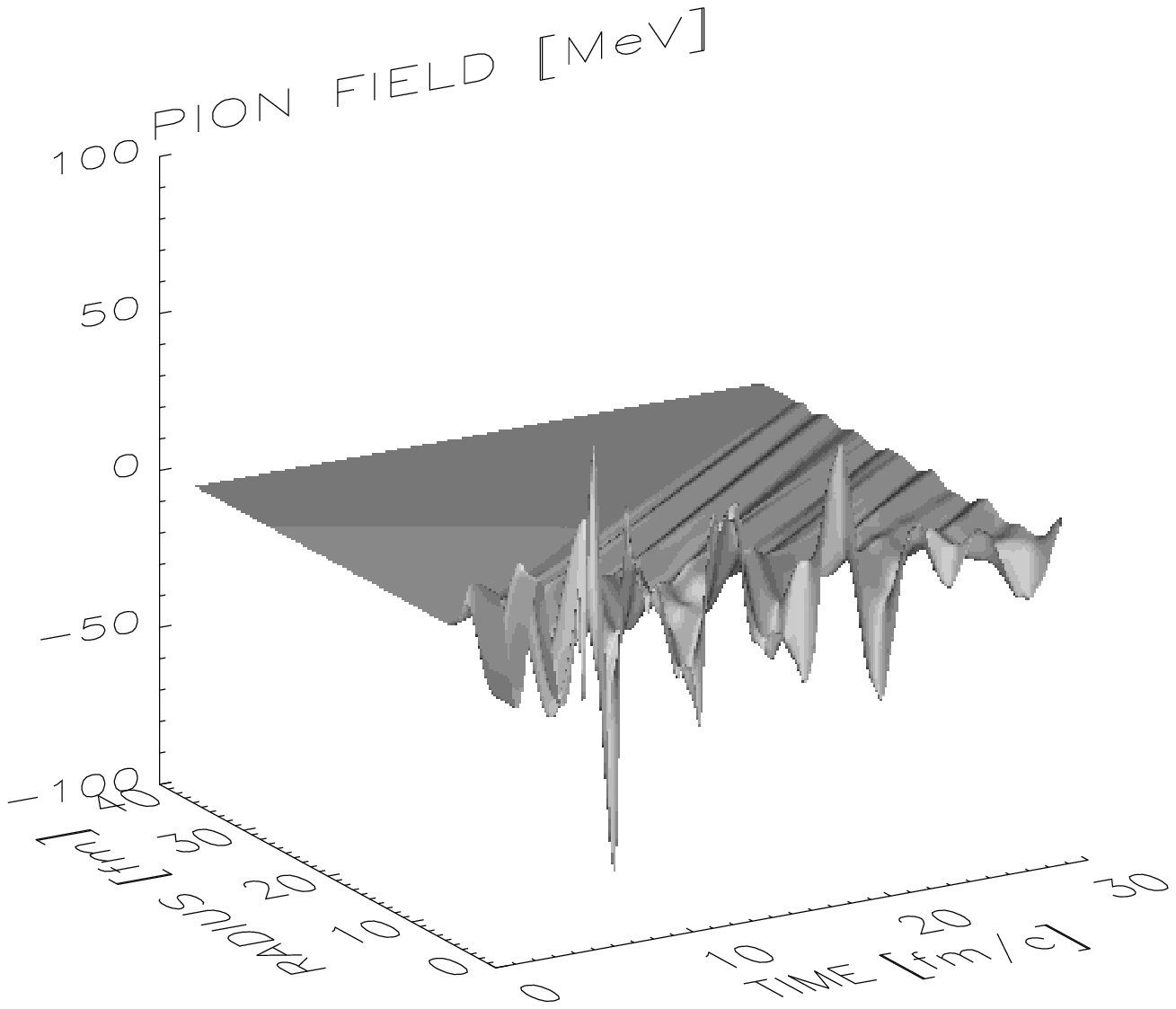}
\epsfxsize=15cm
\epsffile{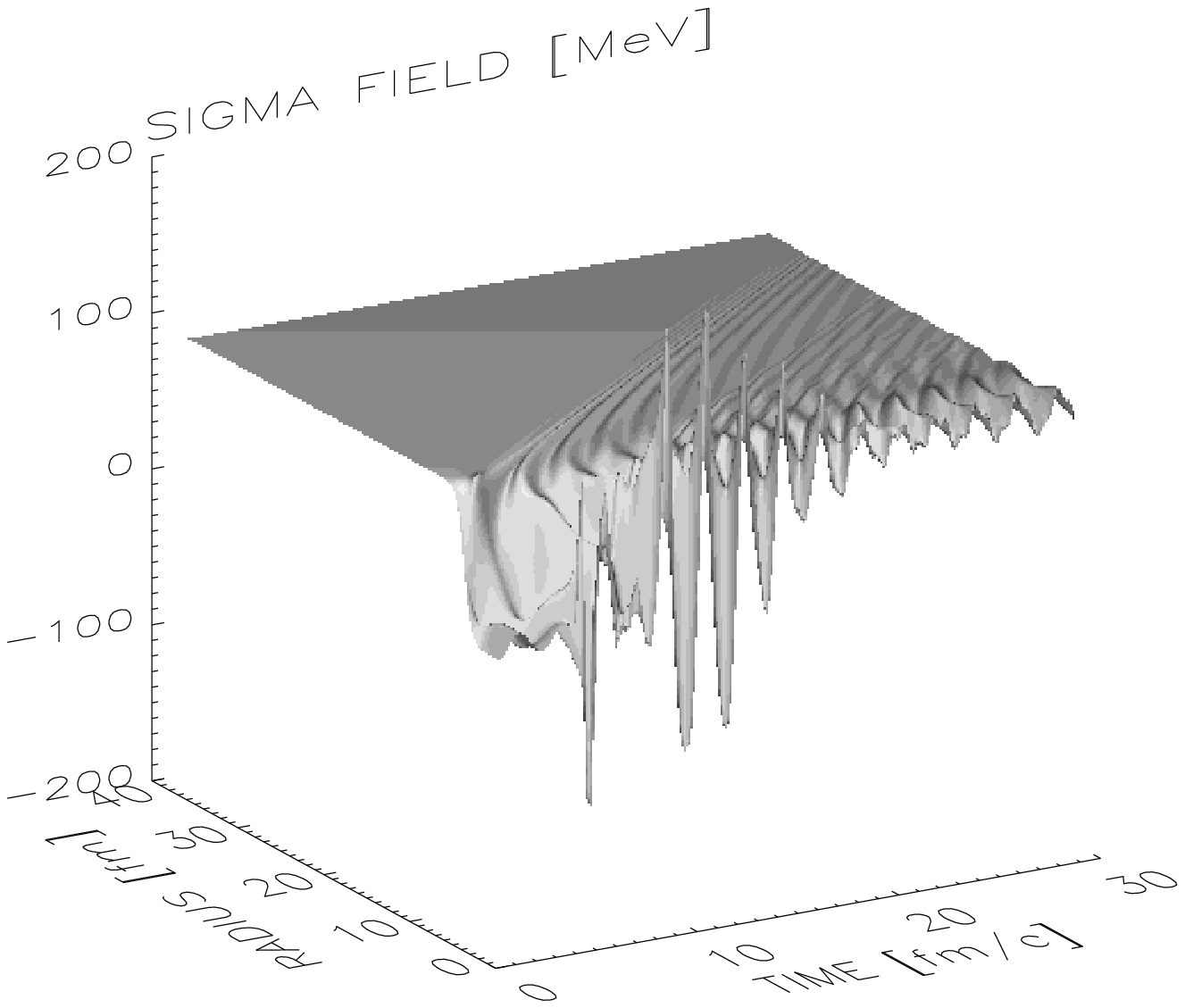}
\newpage
\epsfxsize=15cm
\epsffile{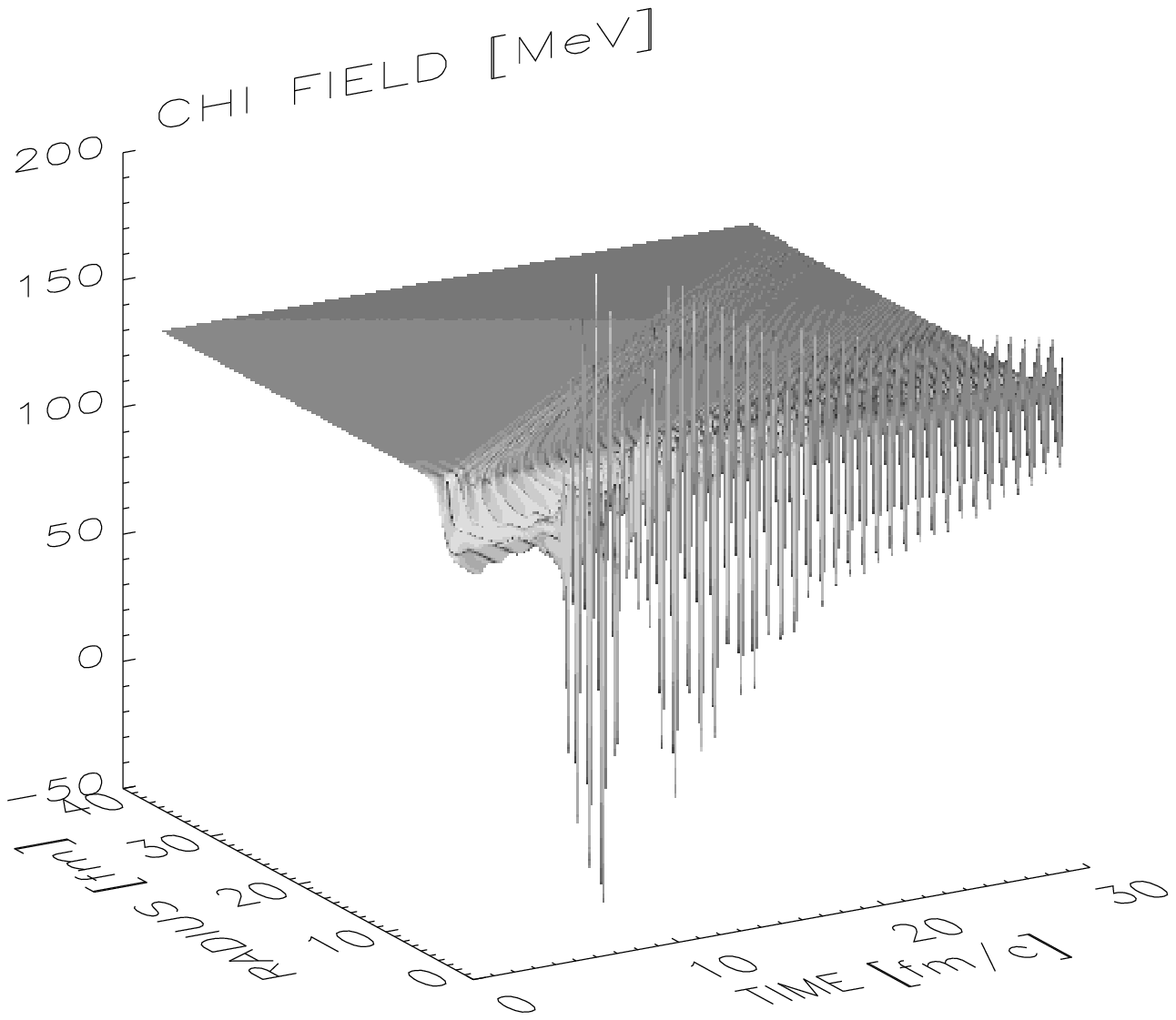}
\caption{The evolution of pion, sigma  and glueball fields in
 the $t-r$ plane for an expanding droplet of initial radius $R=4$ fm.
 The initial values ((case I) of
the fields are: $\sigma=-0.05f_{\pi}$, $\dot{\sigma}=0$,
 $\pi_3=0.2f_{\pi}$, $\dot{\pi_3}=0$,
$\chi=0.79\chi_0$ and $\dot{\chi}=0$.}
\end{center}
\end{figure}
\end{document}